\documentclass[a4paper]{article}

\usepackage[dvipdfmx]{graphicx}
\usepackage{type1cm}
\usepackage{slashbox}
\usepackage{multirow}
\usepackage{lscape}
\usepackage{amsmath}
\usepackage{framed}
\usepackage{setspace}

\usepackage{mathtools}

\begin{document}
\fontsize{10.5pt}{18pt}\selectfont



\fontsize{13pt}{18pt}\selectfont
\begin{center}
{\bf Controllability Analyses on Firm Networks\\ Based on Comprehensive Data\footnote{
This study is conducted as a part of the project
``Price Network and Dynamics of Small and Medium Enterprises''
undertaken at the Research Institute of Economy, Trade and Industry (RIETI).
The authors thank the institute for various means of support.
We thank Hiroshi Yoshikawa, Hideaki Aoyama, Hiroshi Iyetomi, Yuichi Ikeda, Yoshi Fujiwara, Wataru Soma, Yoshiyuki Arata,
and members who attended the internal seminar of RIETI for their helpful comments.
In addition, we thank Yang-Yu Liu for sharing his program codes and giving helpful comments.
We gratefully acknowledge financial support from the Japan Society for the Promotion of Science (No. 15K01217).
}}

\vspace{2ex}

Hiroyasu Inoue\footnote{Graduate School of Simulation Studies, University of Hyogo, 7-1-28 Minatojima-minamimachi, Chuo-ku, Kobe, Hyogo 650-0047, Japan}\\
Graduate School of Simulation Studies, University of Hyogo

\vspace{3ex}

\fontsize{10.5pt}{18pt}\selectfont
Abstract

\end{center}

\fontsize{10.5pt}{18pt}\selectfont
\noindent
Since governments give stimulus to firms and expect the spillover effect by fiscal policies, it is important to know the effectiveness that they can control the economy. To clarify the controllability of the economy, we investigate a firm production network observed exhaustively in Japan and what firms should be directly or indirectly controlled  by using control theory. By control theory, we can classify firms into three different types: (a) firms that should be directly controlled; (b) firms that should be indirectly controlled; (c) neither of them (ordinary). Since there is a direction (supplier and client) in the production network, we can consider controls of two different directions: demand and supply sides. As analyses results, we obtain the following results: (1) Each industry has diverse share of firms that should be controlled directly or indirectly. The configurations of the shares in industries are different between demand- and supply-sides; (2) Advancement of industries, such like, primary industries or other advanced industries, does not show apparent difference in controllability; (3) If we clip a network in descending order of capital size, we do not lose the control effect for both demand- and supply-sides.

\vspace{2ex}


\vspace{5ex}




\vspace{5ex}



\fontsize{10.5pt}{18pt}\selectfont

\newpage


\section{Introduction}

Since the economy is not fully understood regardless of the long history of economic studies,
governments have to intervene it without assurance so that they can lead the economy to a desirable state,
which has been considered necessary.
Concretely, governments give stimulus to firms and prompt the spillover effect
by purchasing goods and services, giving grants to firms, or fine-tuning taxes and so forth.
These are demand-side stimulus.
In addition, there is an opposite way of such policies.
If scarcity of some production is considered a bottleneck of following productions,
governments can help the production by importing goods, and again, giving grants to firms or fine-tuning taxes and so forth.
These are supply-side stimulus.
Those actions taken by governments are called fiscal policies as opposed to financial policies.
Govenments have considered fiscal policies an important determinant of growth \cite{Easterly93,Romp07}.

When governments conduct those demand- or supply-side fiscal policies,
they need an estimation of the effect, especially an estimation of spillover effects.
Generally, the estimation is being based on input-output tables \cite{Leontief36},
which are matrixes of transaction volumes between industries.
It enables us to obtain an estimation of spillover effects caused by stimulus.

The input-output table is a matrix and therefore, all industries are fully connected,
which means every industry has a connection with every other industry.
However, actual production networks consist of one-to-one connections between firms, not the aggregate connections in industries.
What we can imagine is that some firms are not reachable from some firms
and that some firms are susceptible to the spillover effect.

Although the input-output table is a strong tool to predict mass influence of a change in industries to others,
the diversity in reachability and susceptibity in firms cannot be obtained.
Knowing the diversity is important because
if some firms are not reachable, it causes the disparity between firms in the sense of benefit of fiscal policies.
Even if the disparity is not evitable,
it is still imporant to know the outcome of the conducted policies.

This study reveals what firms should be directly/indirectly controlled and
the configuration of them in each industry.
Furthermore,
we analyze how different clipping of the network affects controllability
to reveal controllability with limited budget.
We use exhaustive transaction data of firms in Japan and control theory \cite{Liu11}.

The remainder of this paper is organized as follows.
In Section 2, we introduce the dataset.
Section 3 describes the methodologies that we utilize in analyses.
Section 4 presents the results.
Finally, Section 5 concludes.

\section{Data}

We use datasets, TSR Company Information Database and TSR Company Linkage Database,
collected by Tokyo Shoko Research (TSR),
one of the major corporate research companies in Japan.
The datasets are provided by the Research Institute of Economy, Trade and Industry (RIETI).
In particular, we use the dataset collected in 2012.
As necessary information for our study,
we use identification, capital, industry type, suppliers and clients.
We construct an entire network of firms based on suppliers and clients.
Note that there are up to 24 suppliers and up to 24 clients for each firm in the data.
It may be considered that the constraint limits the number of links for each node.
However, a node can be suppliers of other nodes without limitation,
as long as those clients designate the node as a supplier of firms,
and vice versa.
Therefore, the numbers of suppliers or clients are not limited to 24.
The number of firms, that is, nodes, is 1,109,549.
The number of supplier-client ties, that is, links, is 5,106,081.
This network has direction and the direction is utilized in our study.

We split firms based on industries.
The industries are classified by the Japan Standard Industrial Classification \cite{Ministry13}.
We mainly use the division levels that have 20 classifications.
However, we make alterations to the classification.
Since the classifications ``S: Government, except elsewhere classified'' and ``T: Industries unable to classify''
are less important in our study, we omit them.
In addition, we separate ``I: Whole sale and retail trade'' into wholesale and retail.
The difference of the divisions is not negligible in our study because controls from outside, such as fiscal policies,
often occur in retail.
Therefore, the division level in our study after alterations shows 19 industries.

\begin{figure}[h]
\begin{center}
\includegraphics[width=15cm]{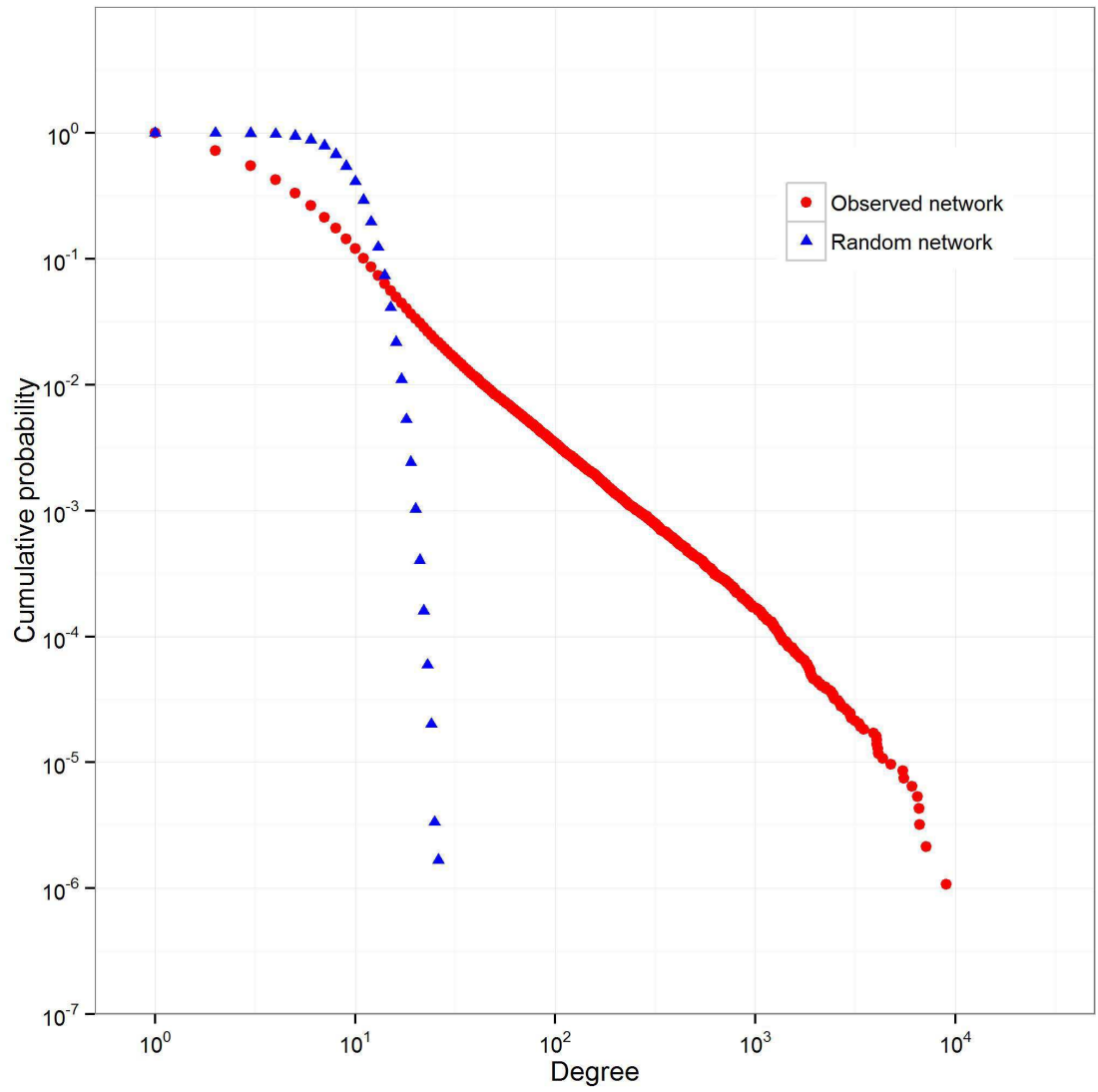}
\caption{Comparison of degree distributions between random network and observed network:
Horizontal axis shows degree and vertical axis shows cumulative probability.
Blue plots indicate random network.
Red ones indicate observed network.}
\label{fig:degreeRandOb}
\end{center}
\end{figure}

Figure \ref{fig:degreeRandOb} shows the degree distribution of the observed network.
The red plots are the distribution of the observed network.
An important point is that the distribution is fat-tailed,
which means the distribution does not decay super-linearly.
It seems that we can fit plots to a line $P\propto k^{-\lambda}$, where $P$ is the cumulative probability,
$k$ is the degree, and $\lambda$ is a positive constant.
If a probability distribution or a cumulative probability distribution
can be fitted to a line, it is said that the distribution is a power-law distribution.
A network with a power-law distribution is often called a scale-free network.
If the degree distribution is the normal distribution,
the plot is shaped as the blue plots in Figure \ref{fig:degreeRandOb}.
Since normal distribution exponentially decays,
we can observe the blue plots decrease super-linearly on the log-log plot.
How to create the random network is explained in Section \ref{cha:results}.
The reason we should compare the observed network with the random network is that
we conduct statistic tests in Section \ref{cha:results}.

\begin{center}
[Figure \ref{fig:degreeRandOb} here]
\end{center}

\section{Methodology}

\label{cha:Methodology}

We use control theory \cite{Kalman63,Luenberger79,Slotine91,Liu11}
to know which nodes are pivotal to control a network.
Here, control means that through directly controlled nodes, other nodes are controlled indirectly.
Control theory tells us whether an entire network is directly/indirectly controllable with a given set of nodes that are controlled directly.
``Controllable'' means that an arbitrary state of a network can lead to any desired state.
We do not have assumption as to what is the ``state'' of firms in this study.
For example, the state can be sales or capitals.
However, as we will explain in this section,
we only consider whether firm A can affect firm B or not
and do not consider volume of trades between them.

Controllability fits with the motivation for fiscal policies.
This is because governments provide stimulus to a certain set of firms
and they try to control firms indirectly.

Control theory can be described formally as follows.
A link indicates that there is a relationship in which one node affects another.
In addition, we can assume that a network receives stimulus from the outside and propagate the stimulus through a given relationship.
Based on this setup, we simply consider the following equation to depict the system.
\[\frac{\mbox{d}{\bf x}(t)}{\mbox{d}t}=A{\bf x}(t)+B{\bf u}(t),\]
where the vector ${\bf x}(t)=(x_{1}(t),\dots,x_{N}(t))^{\mbox{T}}$ is a state of $N$ nodes at time $t$,
the $N\times N$ matrix $A$ is a diagram of links,
the vector ${\bf u}(t)=(u_{1}(t),\dots,u_{M}(t))^{\mbox{T}}$ signifies the strength of outside controllers,
and the matrix $B$ is the $N\times M$ matrix ($M\leq N$) that indicates which drivers
(nodes that take stimulus from outside) are connected to outside controllers.
The system depicted by the equation is controllable
if the following $N\times NM$ matrix
\[C=(B, AB, A^2B,\dots, A^{N-1}B)\]
has full rank. That means
\[\mbox{rank}(C)=N.\]

\begin{figure}[h]
\begin{center}
\includegraphics[width=15cm]{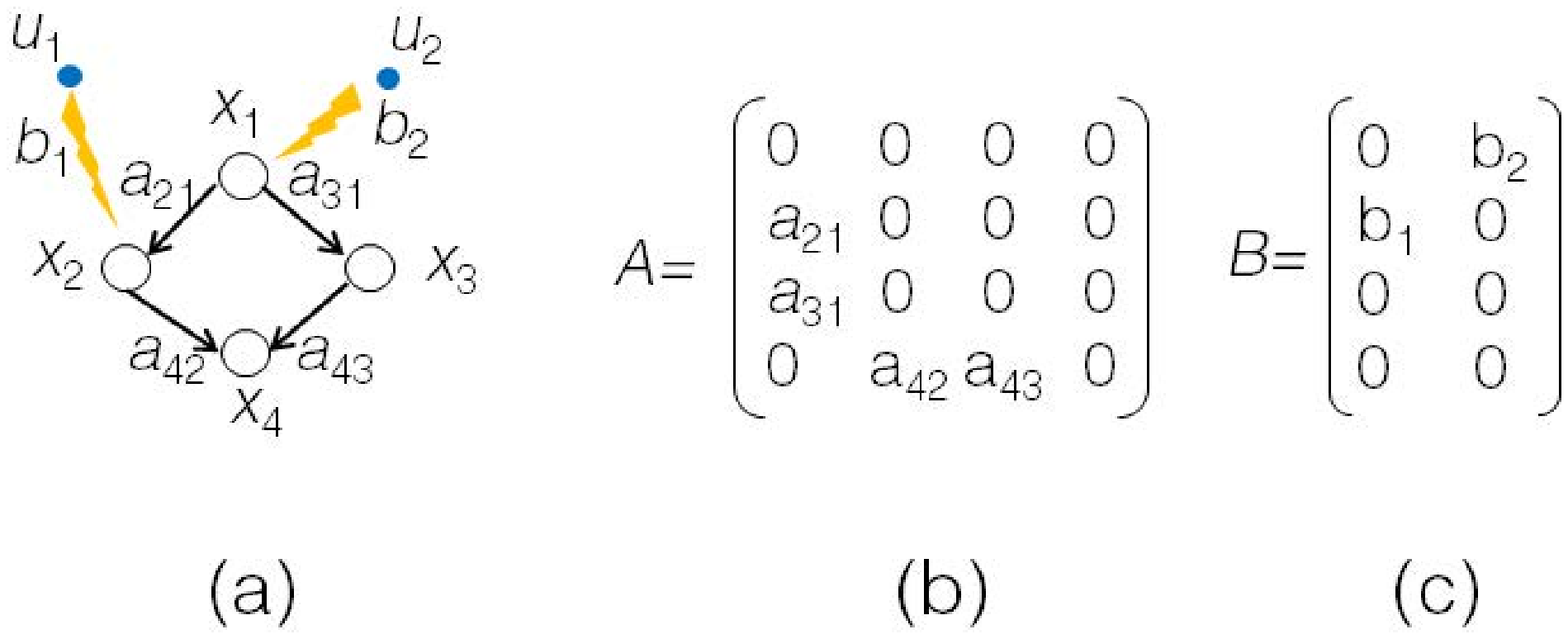}
\caption{Control theory:
(a) Network and two outside controllers.
Network is controllable.
(b) Matrix of network diagram. It corresponds to panel (a).
(c) Matrix of controller-driver diagram. It also corresponds to panel (a).}
\label{fig:schemeControllability}
\end{center}
\end{figure}

Figure \ref{fig:schemeControllability} shows an example of a simple system.
There is a network with four nodes and four links.
In addition, there are two outside controllers.
The matrixes $A$ and $B$ correspond to the network.
The network is controllable.
There are two outside controllers in the example but one outside controller is in theory enough to control for any network.

\begin{center}
[Figure \ref{fig:schemeControllability} here]
\end{center}

Once a set of drivers is given, we can calculate controllability.
However, if we aim to test all sets of drivers, the calculation time is O($2^N$).
The observed data are $N=1,109,549$ and it is obvious that we cannot test all sets without thinking.
Moreover, since a set of drivers that corresponds to all nodes can obviously control a network,
finding the sets for a minimum number of drivers is also important.
Sets of drivers have multiple configurations, even if they have a minimum number of drivers.
Liu et al. developed an algorithm to effectively obtain all sets of drivers that are controllable \cite{Liu11}.

We return to Figure \ref{fig:schemeControllability} to consider the example.
The nodes $x_1$ and $x_2$ are drivers in panel (a).
However, it is obvious that we can choose $x_1$ and $x_3$ for drivers and those are also a minimum set of drivers.
The variable configuration gives us the following distinctions for nodes:
(1) necessary driver: a node that is always chosen as a driver in any configuration of drivers;
(2) necessary follower: a node that is never chosen as a driver in any configuration of drivers;
and (3) ordinary: a node that is possibly chosen as a driver.

Required condition for necessary driver is very easy.
If a node does not have a in-coming link (in-degree), the node is a necessary driver.
Though the condition for necessary driver is easy to be understood,
one for necessary followers is not.
First, a necessary follower should have a link from another node.
We call the source-of-link node a parent node and the target-of-link node a child node.
If parent node(s) do not have other child nodes, or if parent node(s) have other child nodes
and the number of child nodes do not excess one of parent nodes,
the original child node is a necessary follower.

Fiscal policy and control theory are compatible.
For example, it is seemingly more efficient not to choose necessary followers as targets of the fiscal policy than ordinaries or necessary drivers,
because we can expect that indirect effects are delivered to necessary followers.

\section{Results}

\label{cha:results}

\subsection{Control by demand}

\label{cha:demand}


We discuss demand-side control,
which means governments let drivers buy more (possively less) products,
and the increase (or decrease) of demand is delivered to suppliers.
Therefore, control direction is from a client to a supplier.
If a firm has a client, the firm can be controlled indirectly.
The demand-side control means that
governments give stimulus to specific firms
on the assumption that the propagation of the stimulus provides benefit to entire economy.

\begin{figure}[h]
\begin{center}
\includegraphics[width=15cm]{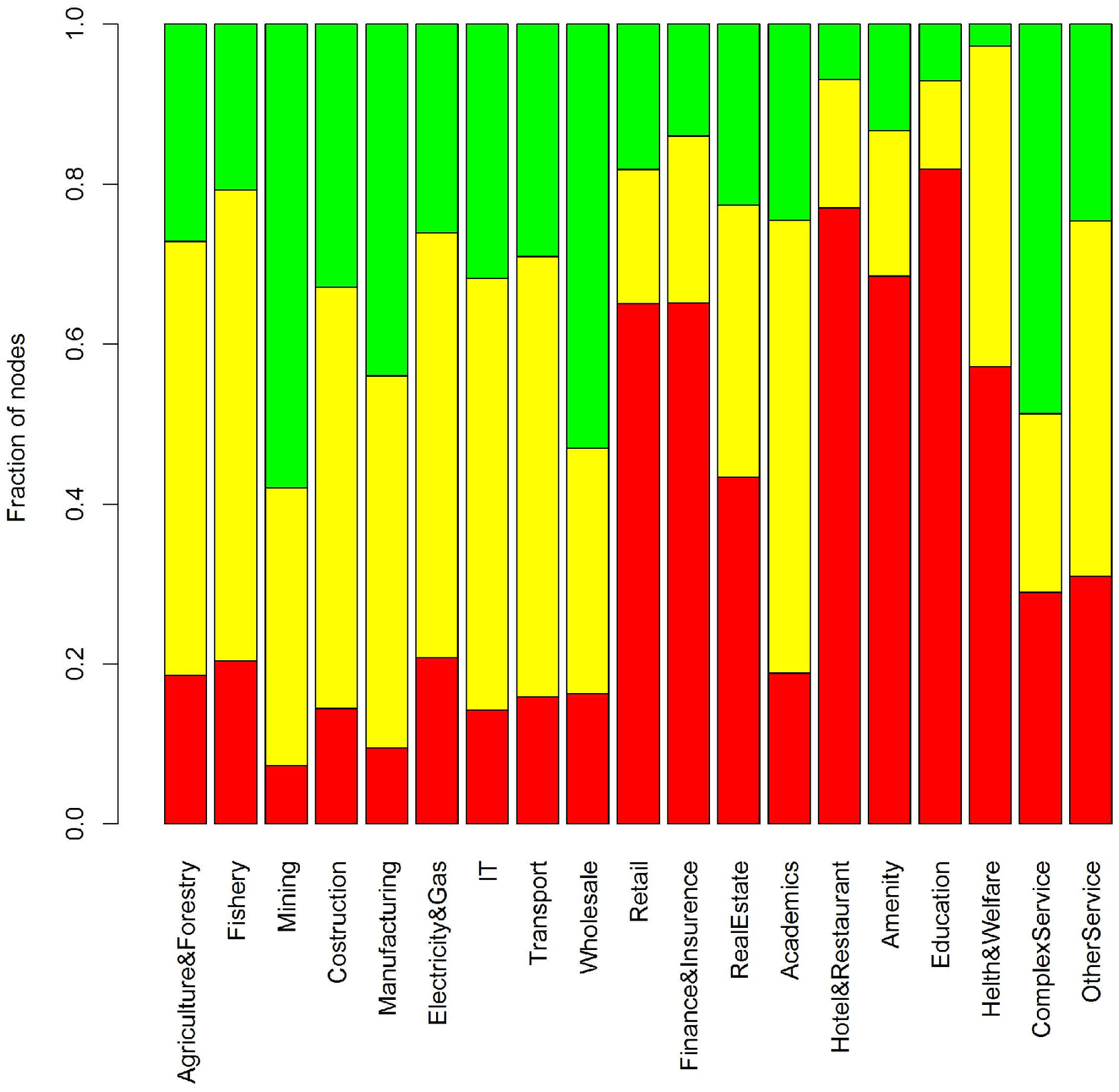}
\caption{Share of node types obtained by control theory (Demand control):
Horizontal axis lists industries.
Vertical axis shows share of necessary followers (green),
ordinary nodes (yellow), and necessary drivers (red).
}
\label{fig:demandControl}
\end{center}
\end{figure}

The results of the calculations are 
given in Figure \ref{fig:demandControl}.
The large share of necessary drivers means
that the industries require direct controls.
Most service industries, including retail, have relatively large shares of necessary drivers.
This is mainly because a lot of firms in those industries do not have clients of firms
and are connected to only final consumers.
In fact, we observe that the degree and the firm size have strong correlation.
Therefore, one of the strong causes of the result seems that small firms are included into the industries.


\begin{center}
[Figure \ref{fig:demandControl} here]
\end{center}

On the other hand, a necessary follower is a firm that should be controlled indirectly.
We can observe mining, manufacturing, and wholesale have relatively large shares of the necessary followers
from Figure \ref{fig:demandControl}.
The large share of necessary followers means
that the industries can be 
controlled indirectly more easily than other industries.
Though necessary drivers tend to be small firms as explained above,
there is no such apparent attribute for necessary followers.
However, as is already mentioned in Section \ref{cha:Methodology},
necessary followers have as many as or more parent nodes.
It can be assumed that an industry with large share of necessary followers has more competitive situations for clients.
This is because there are more clients than suppliers.
If a market of a product is well recognized and a product becomes commodity,
it is natural that there are a lot of clients for the product.




Since firms can be classified into necessary drivers, necessary followers, and ordinary,
they do not have equal sensitivity to the indirect effect.
Therefore, control by governments is inequitably propagated to firms, which causes disparity between firms
and the penetration of the fiscal policy may not be uniform.
If the fiscal policy is based on a trickle-down hypothesis, which means that propagtion is equally propagated,
the implication posed by this study should be considered.

Since it is always difficult to affect the entire economy through fiscal policy,
it is useful to know what happens if we consider a partial network.
Therefore, we apply control theory to clipped networks.
We examine two different ways to clip the observed network: random clipping and capital-order clipping.
In random clipping, a certain amount of nodes are chosen randomly from the observed network
and links that connect the chosen nodes are also reserved.
In capital-order clipping, a certain amount of nodes is chosen by the descending order of the firm's capital size from the entire network
and links between them are reserved.

We clip the observed network with five different fractions:
$2^{-1}$, $2^{-2}$, $2^{-2}$, $2^{-4}$, and $2^{-5}$.
For example, if the fraction is $2^{-1}$, the observed network is clipped into a half size in the sense of the number of nodes.
We obtain unique networks from capital-order clipping
but random clipping is not unique.
Therefore, 
we obtain 10 samples for each fraction for random clipping
and acquire mean and standard deviation.

\begin{figure}[h]
\begin{center}
\includegraphics[width=15cm]{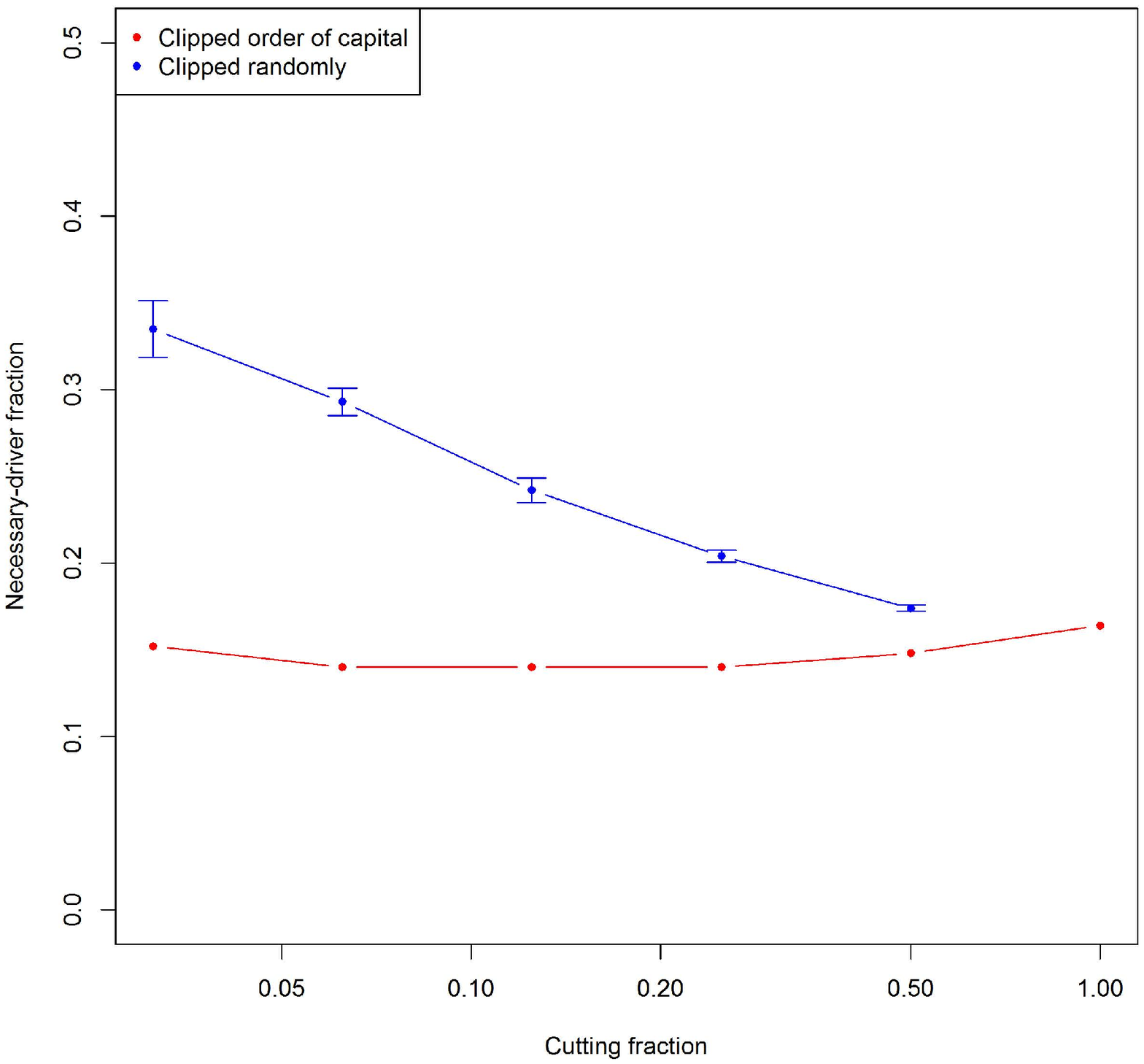}
\caption{
Comparison of necessary drivers between different clippings (Demand control):
Horizontal axis shows fractions of clipping.
Vertical axis shows ratio of necessary drivers.
There are two different measures:
ratio of total capital of necessary drivers
and ratio of total number of necessary drivers.
In addition, there are two different way of clipping:
clipping in order of increasing capital and random clipping.
}
\label{fig:demandClip}
\end{center}
\end{figure}

Figure \ref{fig:demandClip} shows the results.
The horizontal axis indicates the fraction of clipping.
The vertical axis indicates the ratio of the number of the necessary drivers to all nodes.
The bars in the figure show the standard deviations.
We observe that $n_{d}$ decreases or hardly moves for capital-order clipping through different fractions.
On the other hand, random clipping causes an increase of fractions,
which means that the randomly clipped networks are difficult to control.
If we want to partially affect firms,
capital-order clipping seems better than random clipping.
The results of Figure \ref{fig:demandClip} corroborate
the earlier discussion in Figure \ref{fig:demandControl}.
Since small firms tend to be the necessary drivers,
capital-order clipping can avoid those small firms to be chosen.


\begin{center}
[Figure \ref{fig:demandClip} here]
\end{center}

\begin{figure}[h]
\begin{center}
\includegraphics[width=15cm]{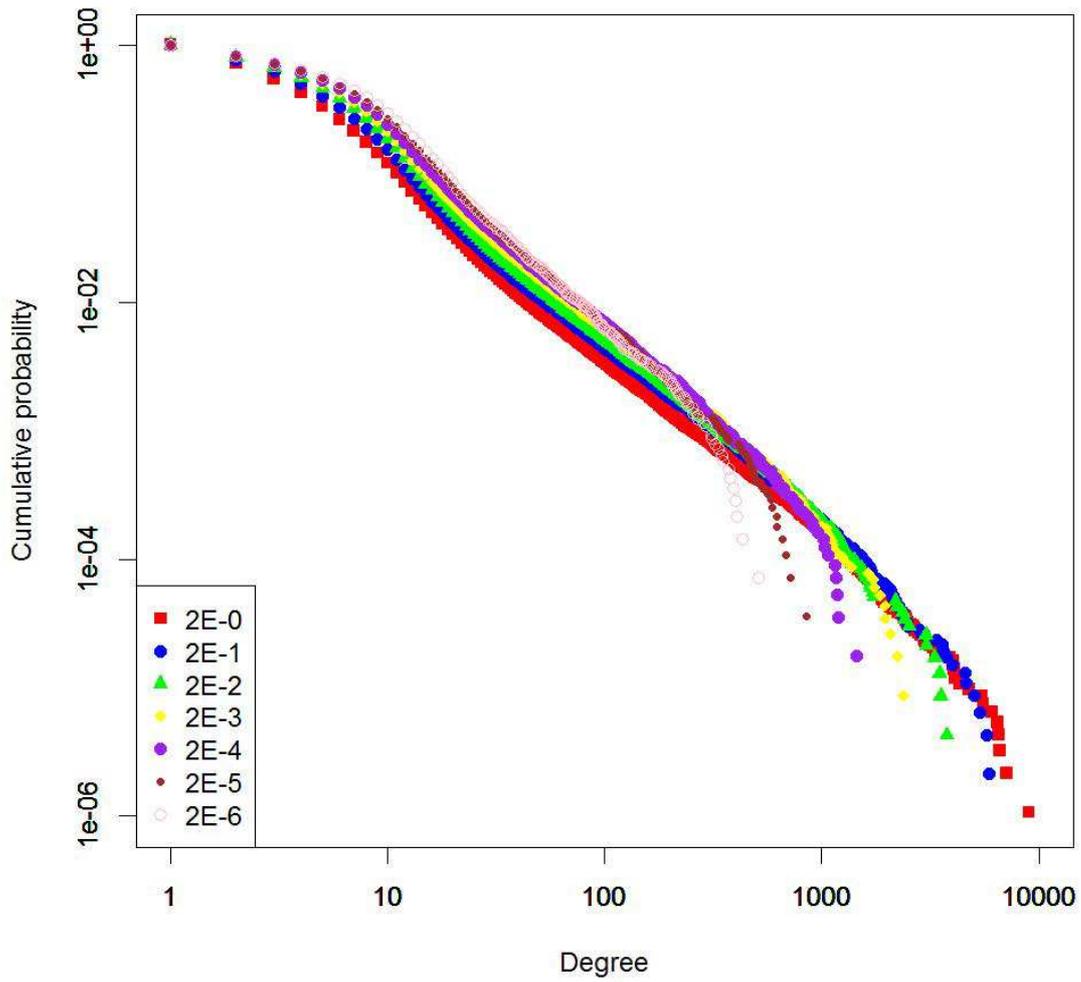}
\caption{Comparison of degree distributions between clipped networks:
Horizontal axis shows degree and vertical axis shows cumulative probability.
Each color and its number correspond to ratio of clipping in order of increasing capital.}
\label{fig:clipDegree}
\end{center}
\end{figure}

Here, we discuss more detail of the results.
For scale-free networks, it is analytically shown that
ratio of drivers depends only on the exponent of the degree distribution and the average of the degree \cite{Liu11}.
(Note that the necessary driver is a part of the driver.)
The equation is
\[n_{d}\approx \exp [-\frac{1}{2}(1-\frac{1}{\gamma-1})\langle k\rangle].\]
Figure \ref{fig:clipDegree} shows the degree distributions of the capital-order clipped networks.
We observe that
all networks are scale-free networks
and the capital-order clipping retains the shapes, which means
the exponents seem to be a constant.
Since small firms roughly have small degrees,
they are cut first in capital-order clipping.
As a result, the remaining firms are densely connected.
Actually, the mean degree $\langle k\rangle$ for the fractions
$2^{0}$, $2^{-1}$, $2^{-2}$, $2^{-2}$, $2^{-4}$, and $2^{-5}$ are
6.00, 6.98, 7.79, 8.42, 8.99 and 9.20.
Therefore, the mean degrees increase as the fraction becomes smaller.
The equation shows that a large mean degree should result in a small ratio of drivers $n_{d}$.
On the other hand, we observe the ratio of the necessary driver for capital-order clipping
is almost constant in Figure \ref{fig:clipDegree}.
Since te necessary drivers are part of drivers,
the ratio of the necessary drivers to drivers increases for small networks.
We can interpret this result as meaning that
capital-order clipping adds importance of necessary drivers to control a network.


\begin{center}
[Figure \ref{fig:clipDegree} here]
\end{center}



\subsection{Control by supply}

We discuss supply-side control,
which means governments let drivers sell more (possibly less) products,
and the change of supply is delivered to suppliers.
Therefore, control direction is from a supplier to a client.
If a firm has a supplier, the firm can be controlled indirectly.
The example of the supply-side control is that
governments help stopping firms with essential goods
because the paucity may stop a lot of other firms' production.
Helping stopping firms can help other downstream firms and it can be said governments indirectly control the downstream firms.

\begin{figure}[h]
\begin{center}
\includegraphics[width=15cm]{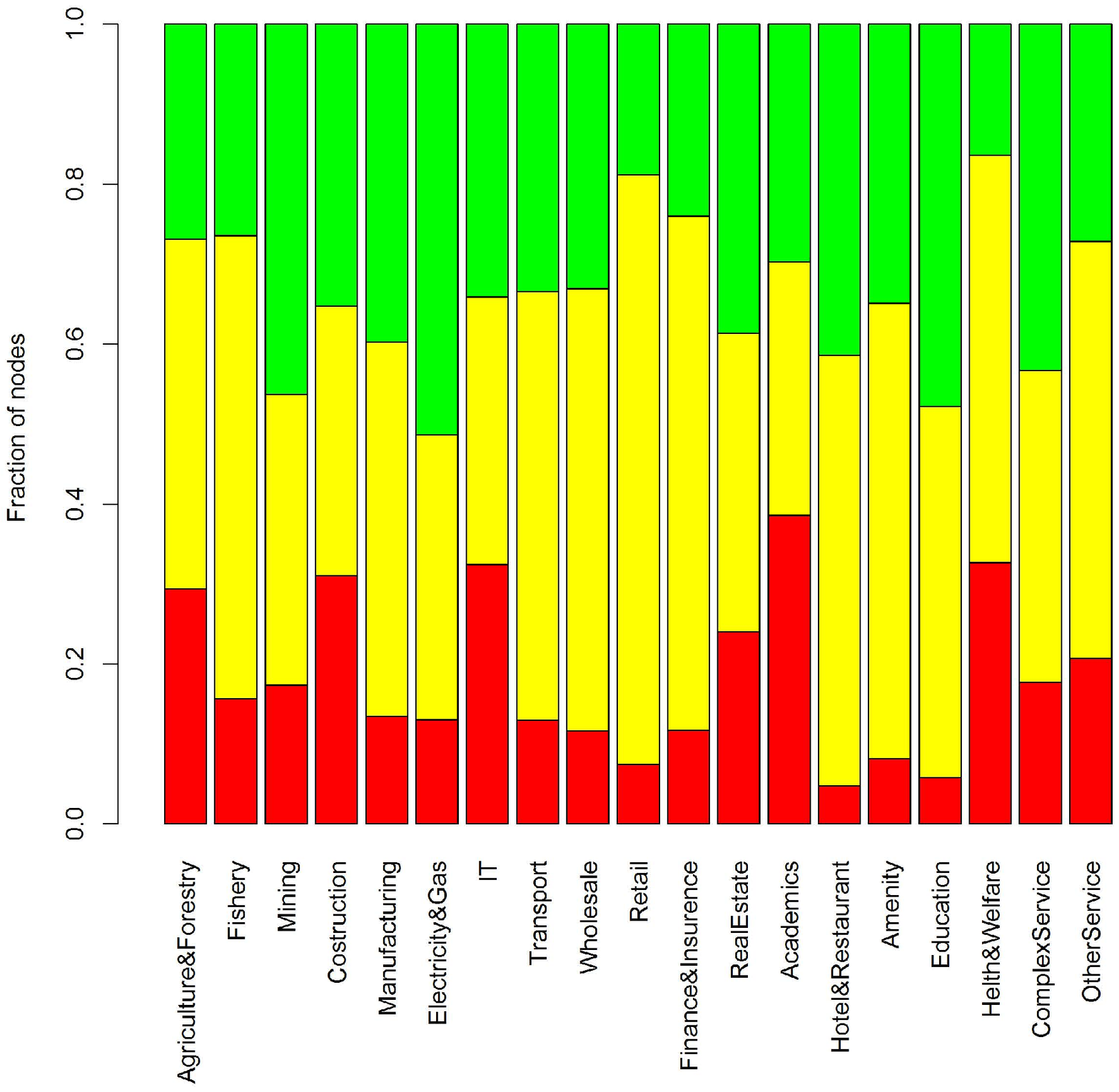}
\caption{Share of node types obtained by control theory (Supply control):
Horizontal axis lists industries.
Vertical axis shows share of necessary followers (green),
ordinary nodes (yellow), and necessary drivers (red).
}
\label{fig:supplyControl}
\end{center}
\end{figure}

Figure \ref{fig:supplyControl} shows
the results of the calculations.
The meaning of the figure is the same as Figure \ref{fig:demandControl}.
Supply-side control is also inequitably propagated to firms, which causes disparity between firms.
This means that the penetration of the fiscal policy may not be uniform.

The large share of necessary drivers means
that the industries require direct controls.
Agriculture \& forestry, construction, IT, academics, and health \& welfare have relatively large shares of necessary drivers.
We note that the industries are different from the result of the demand-side control.
The necessary drivers in supply-side control means that they have no suppliers.
Therefore, first of all, 
the necessary drivers are relatively small firms as is mentioned in the demand-side control.
Moreover, having no supplier is not natural because any firm normally requires some supply.
The transaction data we use in this study records suppliers/clients considered important by each firm.
Hence, there must be ignored trades for all nodes.
This neglect is natural.
For example, every firms buy stationery but we do not call them trades.
Therefore, having no supplier does not necessarily mean that the firm does not buy anything.
Looking back to the industries with a lot of necessary drivers,
the feature of not having suppliers is understandable for
agriculture \& forestry, construction, IT, academics, and health \& welfare.
However, fishery and mining, that belong to the primary industry, could be such an industry without suppliers but they are not.
It seems that those industries in Japan consist of large firms and they tend to have suppliers.

\begin{center}
[Figure \ref{fig:supplyControl} here]
\end{center}

Next, we discuss necessary followers in supply-side control.
An industry with a lot of necessary followers can be
controlled indirectly more easily than other industries.
It was already mentioned that necessary followers 
have as many parent nodes as them or more parent nodes than them.
Necessary followers have 
in supply-side control mean having a lot of suppliers.
We can observe
mining, electricity \& gas, and education have the large shares of the necessary followers
from Figure \ref{fig:supplyControl}.
One possible interpretation of being necessary followers in supply-side control is that
clients are stronger than suppliers
because necessary followers share their parent nodes and there are more parent nodes than them.

\begin{figure}[h]
\begin{center}
\includegraphics[width=15cm]{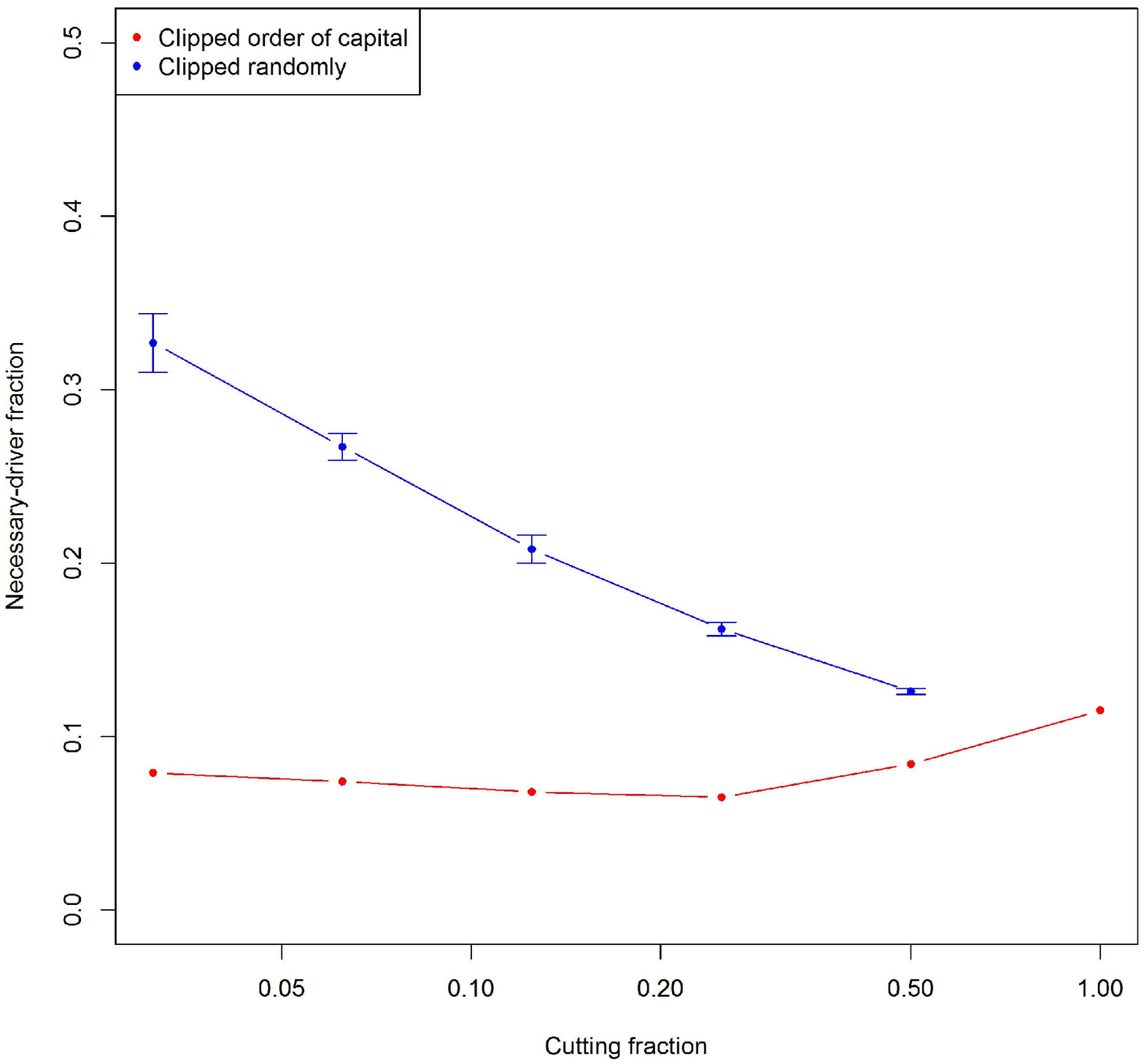}
\caption{
Comparison of necessary drivers between different clippings (Supply control):
Horizontal axis shows fractions of clipping.
Vertical axis shows ratio of necessary drivers.
There are two different measures:
ratio of total capital of necessary drivers
and ratio of total number of necessary drivers.
In addition, there are two different way of clipping:
clipping in order of increasing capital and random clipping.
}
\label{fig:supplyClip}
\end{center}
\end{figure}

We analyze clipped networks by the supply-side control
as is already analyzed by the demand-side control.
Figure \ref{fig:supplyClip} shows the results.
The meaning of Figure \ref{fig:supplyClip} is the same as Figure \ref{fig:demandClip}.
As is the same result as the demand-side control, we observe that
fractions of necessary drivers decrease or hardly move for capital-order clipping through different fractions
and random clipping causes an increase of it.
Therefore, this result means, as is already obtained in Section \ref{cha:demand}, that
capital-order clipping adds importance of necessary drivers to control a network.

\begin{center}
[Figure \ref{fig:supplyClip} here]
\end{center}

\section{Conclusion}

This paper analyzed the controllability of the transaction networks in Japan based on exhaustively collected data.
By using control theory,
we were able to classify firms based on the need to control them.
The result showed that firms are clearly classified into groups based on how they should be controlled
and that the fractions of the classifications in different industries are diverse.
In demand-side control,
most service industries have large shares of necessary drivers
and mining, manufacturing, and wholesale have large shares of necessary followers.
In supply-side control,
agriculture \& forestry, construction, IT, academics, and health \& welfare have large share of necessary drivers
and mining, electricity \& gas, and education have large shares of the necessary followers.
If we clip a network by capital order, we can effectively control the network by necessary drivers in both demand- and supply-side control.

The control theory we adopted in this study cannot incorporate link weight.
However, we may realize it by copying a node that has the same links according to the weight.
This should be studied in the future work.

\bibliographystyle{unsrt}
\bibliography{paper}

\begin{thebibliography}{1}

\bibitem{Easterly93}
W.~Easterly and S.~Rebelo.
\newblock {Fiscal policy and economic growth}.
\newblock {\em Journal of Monetary Economics}, 32:417--458, 1993.

\bibitem{Romp07}
W.~Romp and J.~{de Haan}.
\newblock {Public Capital and Economic Growth: A Critical Survey}.
\newblock {\em Perspektiven der Wirtschaftspolitik}, 8(S1):6--52, 2007.

\bibitem{Leontief36}
{W.W.} Leontief.
\newblock {Quantitative Input and Output Relations in the Economic Systems of
  the United States}.
\newblock {\em The Review of Economics and Statistics}, 18(3):105--125, 1936.

\bibitem{Liu11}
Y.~Liu, {J.J.} Slotine, and {A.L.} Barab{\'a}si.
\newblock Controllability of complex networks.
\newblock {\em Nature}, 473(7346):167--173, 2011.

\bibitem{Ministry13}
Ministry of~Internal~Affairs and Communications.
\newblock The {Japan} standard industrial classification ({JSIC}) summary of
  development of the {JSIC} and its eleventh revision, 2013.

\bibitem{Kalman63}
{R.E.} Kalman.
\newblock Mathematical description of linear dynamical systems.
\newblock {\em Journal of the Society for Industrial \& Applied Mathematics,
  Series A: Control}, 1(2):152--192, 1963.

\bibitem{Luenberger79}
D.~Luenberger.
\newblock {\em Introduction to dynamic systems: theory, models, and
  applications}.
\newblock Wiley, 1979.

\bibitem{Slotine91}
{J.J.E.} Slotine and W.~Li.
\newblock {\em Applied nonlinear control}.
\newblock Prentice-hall Englewood Cliffs, NJ, 1991.

\end{thebibliography}

\clearpage

\clearpage

\clearpage




\clearpage

\clearpage

\clearpage

\clearpage

\clearpage

\end{document}